\documentstyle[11pt,newpasp,twoside,epsf]{article}
\def\edcomment#1{\iffalse\marginpar{\raggedright\sl#1\/}\else\relax\fi}
\reversemarginpar

\begin{document}
\title{The clustering of X-ray and sub-mm sources}
 \author{Omar Almaini}
\affil{Institute for Astronomy, 
University of Edinburgh, Royal Observatory, Blackford
Hill, Edinburgh EH9 3HJ, UK}
\begin{abstract}

It is becoming clear that luminous extragalactic X-ray and sub-mm
sources are essentially distinct populations. Thus, if high redshift
sub-mm sources represent massive spheroids in formation, there must be
a time lag between the major epoch of star-formation and the
appearance of a visible quasar. Despite this distinction, I find
tentative evidence for a puzzling angular cross-correlation between
X-ray sources and bright sub-mm sources in two independent fields.  If
this signal is due to large-scale structure it would argue for a low
redshift ($z<2$) for many of the SCUBA sources.  Alternatively, I
suggest that the effect may be enhanced by gravitational lensing.  The
exceptionally steep slope of the bright sub-mm counts makes this
population particularly prone to even moderate lensing bias.  An
apparent correlation may therefore be produced if X-ray sources trace
the intervening large scale structure.

\end{abstract}

\section{Introduction}

Our understanding of the high-redshift Universe changed dramatically
with the advent of the SCUBA array at the James Clerk Maxwell
Telescope. It appears that a significant (perhaps dominant) fraction
of the star-formation in the high-redshift Universe ($z>2$) took place
in highly luminous, dust-enshrouded galaxies (Smail et al. 1997,
Hughes et al.  1998, Barger et al. 1998, Eales et al. 1999).  The
discovery of this population was heralded by many as the discovery of
the major epoch of dust-enshrouded spheroid formation (Lilly et
al. 1999, Dunlop 2001, Granato et al. 2001).

On a similar timescale, it has become clear that essentially every
massive galaxy in the local Universe hosts a supermassive black hole
(Kormendy \& Richstone 1995, Magorrian et al. 1998).  In particular,
the remarkable relationship between the black hole mass and the
spheroidal velocity dispersion suggests a possible link between an
early epoch of quasar activity and the formation of the spheroid
(Gebhardt et al. 2000, Ferrarese \& Merritt 2000).

Observationally, however, it would appear that only a small fraction
of SCUBA sources display any signs of powerful AGN activity (Fabian et
al.\ 2000; Severgnini et al.\ 2000; Hornschemeier et al. 2000).  This
argues for a time-lag between these two stages of evolution. I present
further evidence to support this scenario, combined with a potential
detection of clustering between X-ray and sub-mm sources. Possible
explanations are briefly discussed.

\section{The ELAIS N2 field}

The 8mJy SCUBA survey is the largest extragalactic sub-mm survey
undertaken to date, covering $260$ arcmin$^2$ in two regions of sky
(ELAIS N2 and the Lockman Hole East) to a typical rms noise level of
$\sigma=2.5$~mJy at $850~\mu$m (Scott et al. 2002). Chandra
observations of the N2 field were presented in Almaini et
al. (2002). The main results can be summarised as follows:

(1) Only $1/17$ SCUBA sources are detected by Chandra.  For the
remaining SCUBA sources our X-ray upper limits allow us to strongly
rule out an AGN SED, even with a large absorbing column, unless the
central engine is completely obscured by Compton-thick material.

(2) We find evidence for angular clustering between the X-ray and
sub-mm populations, with a significance of $3.5-4\sigma$ within $100$
arcsec.  This surprising result appears to suggest that a large subset
of the two populations are tracing the same large-scale structure (see
Figure 1a).

\section{The HDF flanking fields} 

The clustering seen in the ELAIS N2 field clearly requires
confirmation in an independent field.  Borys et al. (2001) have
recently obtained a contiguous, shallow SCUBA scanmap covering an
$11\times11$ arcmin region centred on the HDF. They find 12 sources
brighter than $\sim $10mJy with a significance above $3.5\sigma$. For
the X-ray population I use the recently published catalogue from the
1Ms Chandra observation of this region (Brandt et al. 2001). To ensure
a uniform X-ray coverage across the SCUBA map I choose only the X-ray
sources with a flux above $5\times 10^{-16}~$erg$\,$s$^{-1}$cm$^{-2}$
($0.5-8.0$~keV). Cross-correlating these populations, I obtain a
$2.4\sigma$ excess of pairs within $100$ arcsec.

Barger et al. (2000) have surveyed this region to greater depth with
SCUBA by targeting optically-faint radio sources. This is arguably the
most efficient way to find SCUBA sources, although somewhat incomplete
and potentially biased against very high redshift objects ($z>3$).
Nevertheless, I supplement the Borys et al. catalogue by adding 6
additional sources from Barger et al. to reach a `flux limit' of 7mJy
(comparable with ELAIS N2). With potential biases in mind, we note
that the addition of these sources boosts the significance of the
cross-correlation signal to $2.9\sigma$ within $100$ arcsec (see
Figure 1b).

\begin{figure}
\plotone{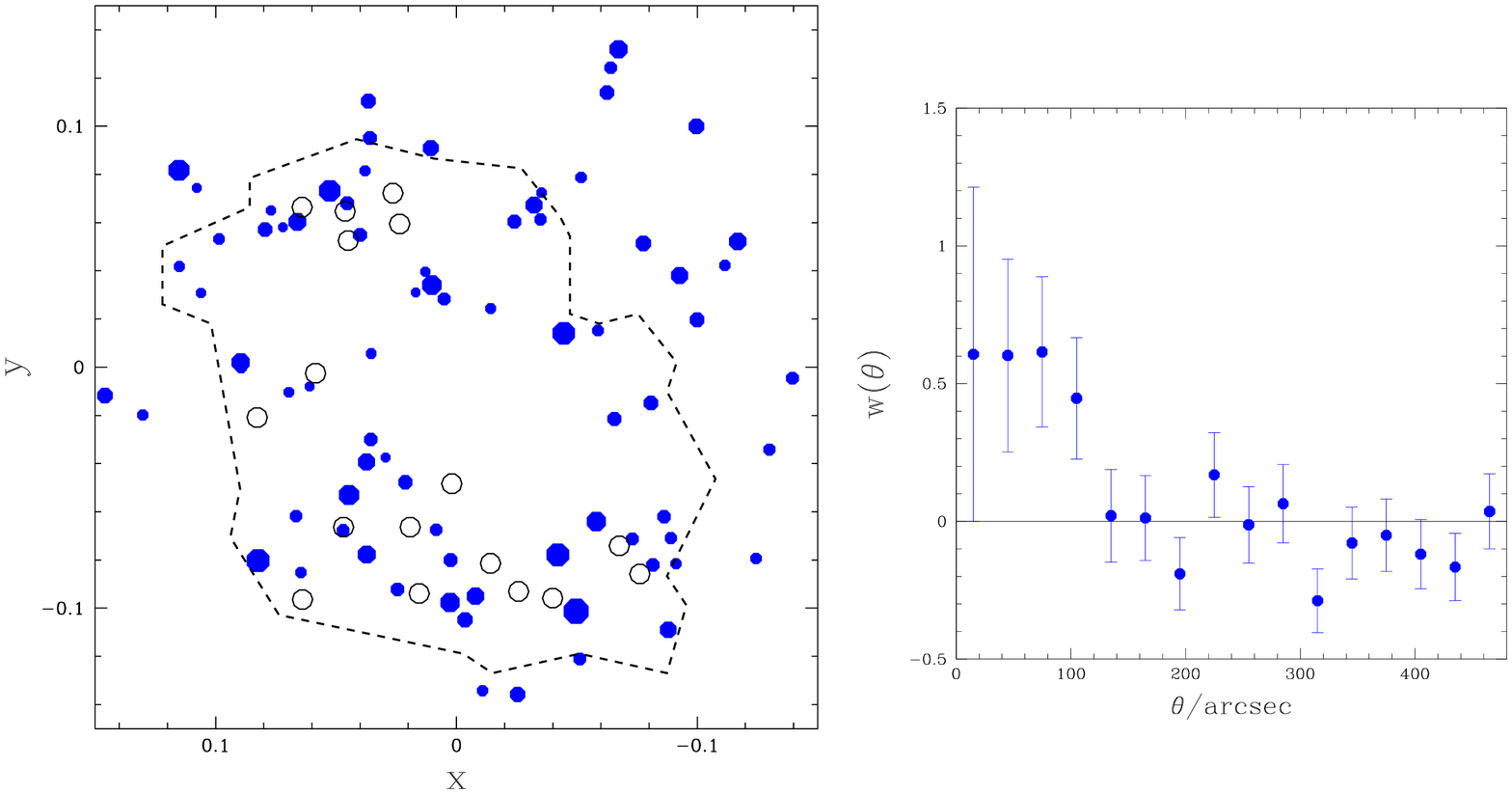} 
\vspace{0.1cm}
\plotone{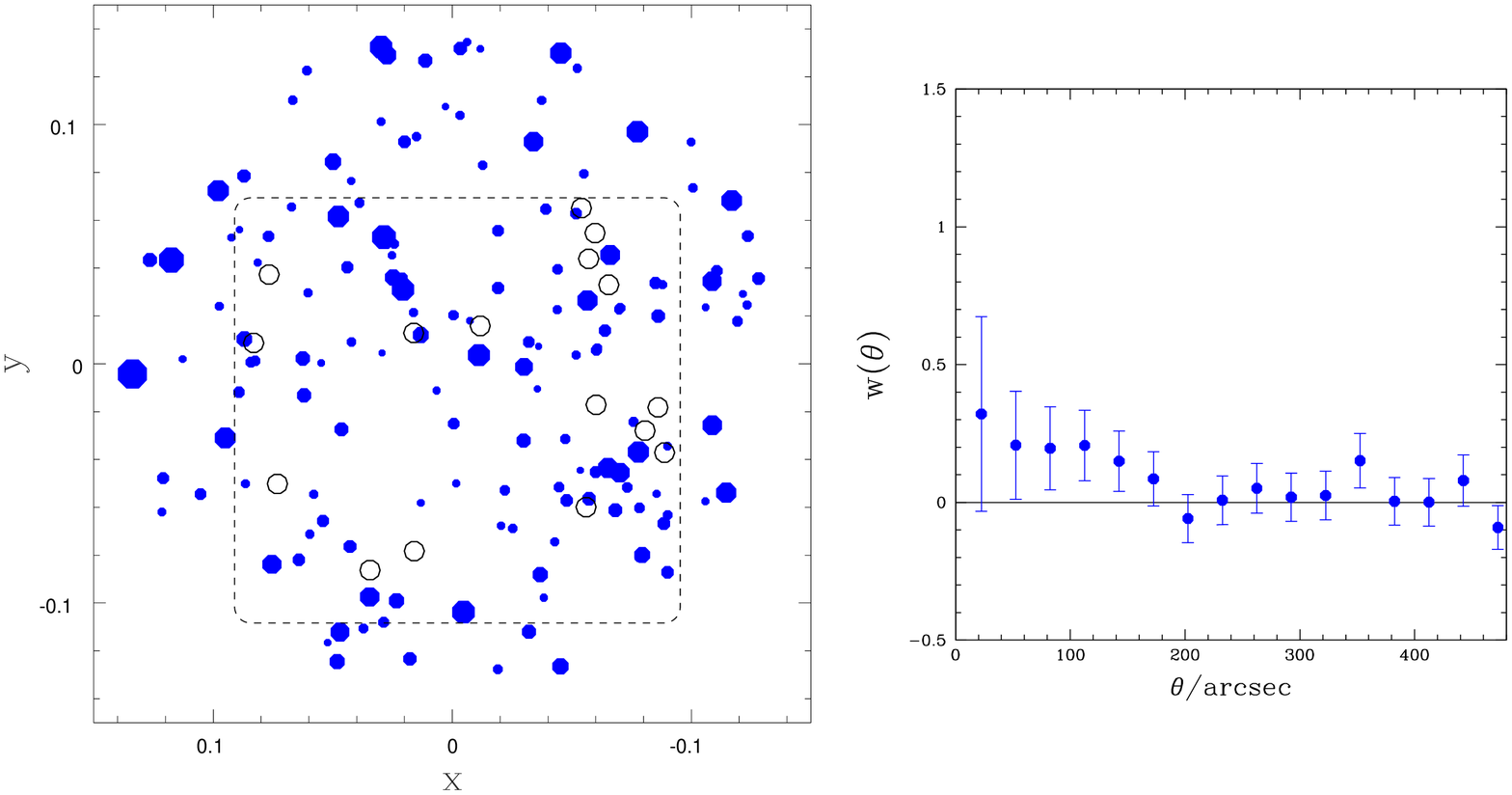}
\caption{Distribution of X-ray (filled points) and SCUBA sources (open
circles) in the ELAIS N2 field (top) and HDF (bottom). The size of the
X-ray points are proportional to the log of their flux. The dashed
regions show the extent of the SCUBA coverage. A statistical
cross-correlation, $w(\theta)$, is shown in each case. }
\end{figure}

\section{Discussion}

It would appear that only a small fraction of SCUBA sources are
detected by Chandra as luminous X-ray sources.  The implication is
that (for a given galaxy) the major episode of star-formation is
essentially distinct from the period of observable quasar
activity. Are the black holes in the remaining SCUBA sources dormant,
still growing or very heavily obscured? This remains a major
unanswered question, which we are currently investigating.

In addition to the low coincidence rate, for two independent fields I
find evidence for clustering between bright SCUBA sources and X-ray
selected AGN. If this is real large-scale structure the redshift
distribution of the X-ray sources would argue for a large fraction of
the SCUBA population lying at $z<2$. I propose an alternative
explanation which invokes gravitational lensing, motivated by the
particularly steep sub-mm source counts.  In the `weak' lensing
regime, for example, where the typical magnification $\mu$ is not
significantly greater than unity, one can readily demonstrate that a
population with cumulative number counts given by a power law of index
$\beta$ will be modified as follows:

\begin{equation}
N'(>S) = \mu ^{\beta -1} N(>S)
\end{equation}

Observed sub-mm number counts have a slope with $\beta \simeq 2.5$
(possibly steepening beyond 8mJy). Foreground large scale structure
(e.g. foreground groups) can readily lead to a magnification of $\mu
\simeq 1.2-1.3$.  Thus in the vicinity of such structure one would
{\em expect} an enhancement in number counts of $30-50$\% which could
easily produce a positive cross-correlation with foreground
populations.

Estimates for the global fraction of lensed sources have been modeled
by Blain et al. (1999) and Perrotta et al. (2001) and predicted to be
only a few per cent. The fields analysed here may therefore be
particularly rich in foreground structure, and this is under
investigation. Alternatively, the lensing cross-sections assumed in
the models could easily be underestimated (due to halo substructure,
for example). In addition, the {\em intrinsic} slope of the bright
sub-mm counts may be turning over more rapidly than previously
assumed, perhaps reflecting a steep turn over in the underlying sub-mm
luminosity function.

\end{document}